\documentclass{article}

\usepackage{arxiv}

\usepackage[utf8]{inputenc} 
\usepackage[T1]{fontenc}    
\usepackage{hyperref}       
\usepackage{url}            
\usepackage{booktabs}       
\usepackage{amsmath}
\usepackage{nicefrac}       
\usepackage{microtype}      
\usepackage{cleveref}       
\usepackage{lipsum}         
\usepackage{graphicx}
\usepackage{natbib}
\usepackage{doi}

\usepackage{multirow}
\usepackage{xcolor}
\hypersetup{
    colorlinks,
    linkcolor={red!50!black},
    citecolor={blue!50!black},
    urlcolor={blue!80!black}
}

\title{On the calibration of neural networks for histological slide-level classification}

\date{}

\usepackage{authblk}

\author[1]{Alexander Kurz}
\author[1]{Hendrik A. Mehrtens}
\author[1]{Tabea-Clara Bucher}
\author[1]{Titus J. Brinker\thanks{Correspondence to \texttt{titus.brinker@dkfz.de}}}
\affil[1]{Division of Digital Biomarkers for Oncology, DKFZ, Heidelberg, Germany}

\renewcommand{\shorttitle}{Slide-Level Calibration in Histopathology}

\hypersetup{
pdftitle={Slide-Level Calibration in Histopathology},
pdfsubject={cs.CV},
pdfauthor={Kurz et al.},
pdfkeywords={histopathology, whole slide image, uncertainty estimation, network calibration},
}

\begin{document}
\maketitle

\begin{abstract}
Deep Neural Networks have shown promising classification performance when predicting certain biomarkers from Whole Slide Images in digital pathology.
However, the calibration of the networks' output probabilities is often not evaluated.
Communicating uncertainty by providing reliable confidence scores is of high relevance in the medical context.
In this work, we compare three neural network architectures that combine feature representations on patch-level to a slide-level prediction with respect to their classification performance and evaluate their calibration.
As slide-level classification task, we choose the prediction of Microsatellite Instability from Colorectal Cancer tissue sections.
We observe that Transformers lead to good results in terms of classification performance and calibration. 
When evaluating the classification performance on a separate dataset, we observe that Transformers generalize best.
The investigation of reliability diagrams provides additional insights to the Expected Calibration Error metric and we observe that especially Transformers push the output probabilities to extreme values, which results in overconfident predictions.

\end{abstract}


\section{Introduction}
\label{sec:introduction}

In Digital Pathology, a common task is to derive a slide-level diagnosis from digitized tissue sections.
Due to the large size of Whole Slide Images (WSIs), specific methods need to be developed, that are able to infer a relevant diagnosis from such large-scale input.
In the past years, Deep Learning (DL)-based systems have shown promising results on predicting certain biomarkers from digitized tissue \cite{kather2019, kiehl2021}.
While these systems typically reach good performance in terms of classification accuracy, a central problem to actually translate these systems into clinical routine is that they are poorly calibrated and provide overconfident predictions \cite{guo2017, minderer2021}.

Within this work, we aim to analyze the calibration of the most widely used modern neural network architectures on the task of Microsatellite Instability (MSI) prediction on Colorectal Cancer (CRC). 
MSI is an important biomarker for CRC patients that allows for specific treatment with immune checkpoint inhibitors \cite{echle2021}.
Several previous works \cite{kather2019, bilal2021, guo2023, wagner2023a} have shown promising results on predicting the MSI status directly from WSIs by using DL-based systems.

The typical workflow for generating a slide-level diagnosis consists of \textbf{a)} Dividing the WSI into small image patches, \textbf{b)} Extracting feature representations for each patch and \textbf{c)} Aggregating the feature representations towards a slide-level prediction.
\cite{ghaffarilaleh2022} provides a comprehensive overview of existing approaches for classifying WSIs.
When comparing and developing new approaches, most works focus on improving the classification performance for the task at hand, such as MSI prediction. We argue that besides improving classification performance, it is important to also evaluate the calibration of the proposed approach.
Especially in the medical context and here for the example of digital pathology, the practitioner needs to receive a reliable estimate of the model's confidence in its prediction.
Calibrated confidence scores are a way to communicate the model's uncertainty \cite{mehrtens2023b}, which is a necessary requirement for medical decision making \cite{begoli2019, kompa2021}.

Within this work, we evaluate the calibration for slide-level MSI status prediction for three representative feature aggregation models.
To reproduce our experiments, our code is available at \url{https://www.github.com/DBO-DKFZ/wsi-calib}.

\section{Methods}
\label{sec:methods}

In this work, we compare three different network architectures for slide-level classification on the task of predicting MSI from colorectal tissue. 

\subsection{Data Preparation}
\label{sec:data-preparation}
To train and evaluate our models, we use data from the following two sources:

From the Molecular and Cellular Oncology (MCO) study \cite{hawkins2014, ward2014, jonnagaddala2016}, we extract 1,313 WSIs of which 175 have label MSI and the other 1,138 are Microsatellite Stable (MSS).

From The Cancer Genome Atlas (TCGA, \url{https://portal.gdc.cancer.gov/}), we access all WSIs from the TCGA-COAD and TCGA-READ studies. 
The MSI labels are assigned following \cite{kather2019}.
After filtering out slides with undesired marker artifacts, we end up with 326 slides of which 60 are MSI and 266 are MSS.

For processing the large-scale images, we first perform tissue detection following the example of \cite{khened2021}.
After identifying the image regions that contain tissue, we store the pixel location of each image patch at the highest available resolution.
We make sure that all included WSIs have a comparable pixel resolution of approximately $0.25 \, \mu\text{m/pixel}$ at  the highest magnification level.

Within our work, we only concern methods that aggregate feature representations of individual patches to infer a slide-level prediction.
Therefore in a following step, we extract a feature representation for each image patch by using a Convolutional Neural Network (CNN).
In the field of deep learning for histopathology, there exist numerous approaches for extracting features from image patches.
In our work, we evaluate feature vectors retrieved from a ResNet-18 \cite{he2016}, which has either been trained on the ImageNet dataset \cite{deng2009} or specifically on histological images \cite{ciga2022}.

\subsection{Network Architectures}
\label{sec:network-architectures}
To this end, we compare three methods for aggregating patch-level feature representations towards a slide-level prediction.
As baseline we choose the CLAM architecture \cite{lu2021}, a method based on the concept of Multiple Instance Learning.
The special property of CLAM is that it combines the slide-level cross-entropy loss with an instance-level clustering loss. 
This leads to the total loss weighted by parameters $c_1$ and $c_2$ as follows:
\begin{align}
    \mathcal{L}_{total} = c_1 \mathcal{L}_{slide} + c_2 \mathcal{L}_{instance}
\end{align}
For the weights we choose $c_1 = 0.7$ and $c_2 = 0.3$ as in the original work.

As second network architecture in our comparison we choose the popular Transformer architecture.
Originating from Natural Language Processing \cite{vaswani2017}, Transformers have also become popular in image processing \cite{dosovitskiy2021}.
Following their successes in natural image processing, Transformers have been applied to histopathology in other previous works \cite{chen2022a, guo2023, wagner2023a}.
While most works also apply a Vision Transformer (ViT) for the feature extraction step from the individual image patches, we use the CNN-based features across methods for comparability.
This way, the patch-level feature representations are viewed as input sequence for a sequence-based classification task.
As model parameters, we choose an embedding dimension of $256$ and stack $4$ Transformer encoder layers, each consisting of a Multi-Head Attention layer with $4$ heads and a feed-forward network with the size of the hidden dimension set to $512$.
Since the computational complexity of the self-attention mechanism scales quadratically with the sequence length, we set a maximum sequence length of $n = 5000$ patches. 
If the number of patches for one slide exceeds this limit, we randomly sample $n$ patches from the slide.

The third architecture we include are Graph Neural Networks (GNNs).
After dividing the WSI into multiple patches, it appears straightforward to build a graph from the individual patches to represent the complete WSI.
Other works have already applied GNNs to different prediction tasks in histopathology, see \cite{ahmedt-aristizabal2022}.
Compared to CNNs, there do not exist established architectures for GNNs, since the graph network architectures depend much on the data.
For constructing the graph representation of one WSI, we connect the feature representation of each image patch (the graph's nodes) with its neighboring patch representations.
For the message passing between the graph nodes, we choose Graph Attention layers \cite{velickovic2018}.
The network we use consists of three sequential Graph Attention layers and the WSI classification is performed by computing a prediction for the whole graph via a pooling layer.

\subsection{Network Calibration and Temperature Scaling}
\label{sec:calibration-and-ts}
The focus of our work is to compare the calibration of the predicted output probabilities of the different feature aggregation methods.
A network is well calibrated if the output probability equals the average accuracy of the predictions.
This means that for the example of an output probability of $80 \%$, the predicted label is expected to be correct in $80 \%$ of the cases.

To measure the calibration over all outputs, a common metric is the Expected Calibration Error (ECE).
To compute the ECE, the output probabilities are divided into $M$ bins, with the number of bins typically being set to $M=10$.
The ECE is then computed as weighted difference between average confidence and average accuracy for each bin as follows:
\begin{align}
    \text{ECE} = \sum_{m=1}^M \frac{|B_m|}{n} \left| \text{conf}(B_m) - \text{acc}(B_m) \right|
\end{align}

Often it is helpful to present reliability diagrams along with the scalar ECE value to gain better insights how the ECE value is computed as done in \cite{guo2017, minderer2021}.
From the reliability diagrams, it can be observed whether the model tends to provide overconfident or underconfident predictions.

In \cite{guo2017} the authors show that a single scaling parameter $T$ can help to improve network calibration.
This parameter is denoted as "Temperature" to scale the output logits of the neural network before applying the softmax function.
The scaled output probability $\hat{q}_i$ is then computed as
\begin{align}
    \hat{q}_i = \max_k \sigma_{SM}(\mathbf{z}_i/T)^{(k)}
\end{align}
where $\mathbf{z}_i$ is the logits vector of the $i$-th sample and $k$ is the number of classes.

\section{Experiments and Results}
\label{sec:results}

For our experiments, we randomly select $20 \%$ of the MCO slides as test data. 
The remaining MCO slides are used to perform $5$-fold cross-validation.
We train all our models for $200$ epochs and select the model with the best balanced accuracy on the validation set for each fold.
During training, we balance the classes on the training dataset and validate on the full validation dataset.
After training, we evaluate the performance of the trained model on the in-distribution (ID) test dataset and also evaluate the performance on the TCGA slides, which we use as out-of-distribution (OOD) data.

\begin{table}[htbp]
    \centering
    \caption{
    Classifier AUROC depending on magnification level and feature extractor.
    The AUROC values are computed on the ID test dataset across five models trained on five different folds.
    }
    \label{tab:resolution_performance}
    \begin{tabular}{lllll}
    \multicolumn{2}{c}{} & \multicolumn{3}{c}{Feature Aggregation} \\\cline{3-5}
    Magnification & Feature Extractor & CLAM & Transformer & GNN \\
    \hline
    \multirow{2}{*}{20x}    & ResNet-18 ImageNet    
    & $0.763 \pm 0.027$ & $\mathbf{0.859 \pm 0.014}$ & $0.877 \pm 0.025$ \\
                            & ResNet-18 Ciga        
    & $\mathbf{0.775 \pm 0.048}$ & $0.858 \pm 0.018$ & $\mathbf{0.885 \pm 0.021}$ \\
    \hline
    \multirow{2}{*}{40x}    & ResNet-18 ImageNet    
    & $0.745 \pm 0.033$ & $0.853 \pm 0.017$ & $0.868 \pm 0.015$ \\
                            & ResNet-18 Ciga        
    & $0.754 \pm 0.029$ & $0.857 \pm 0.020$ & $0.835 \pm 0.020$ \\
    \hline
    \end{tabular}
\end{table}

As mentioned in Section \ref{sec:data-preparation}, we compare the influence of two different feature extractors on the performance of the slide-level classification task.
In \autoref{tab:resolution_performance} we compare the ResNet-18 pre-trained on ImageNet against the one pre-trained on histopathological data (denoted as Ciga) for two different magnification levels (20x and 40x).
We observe that the slide-level MSI classification performance in terms of AUROC is better with the lower 20x magnification level.
The two different feature extractors almost perform on par across the three different aggregation methods.
For the following experiments we will therefore use a magnification level of 20x and we will continue to include both feature extractors.

\begin{figure}[htbp]
    \includegraphics[width=\textwidth]{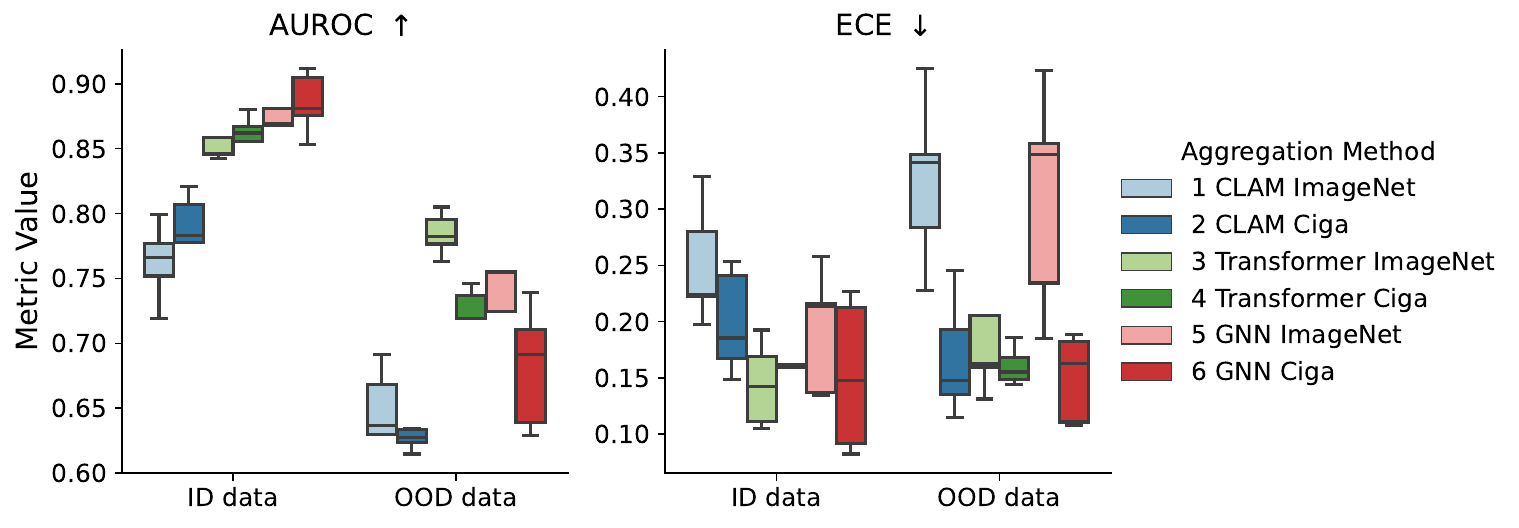}
    \caption{Slide-level classification performance on 20x magnification w.r.t. aggregation method and feature extractor.}
    \label{fig:metrics}
\end{figure}

In \autoref{fig:metrics} we show a barplot to compare classification performance in terms of AUROC and calibration in terms of ECE.
For the ID test set, we see increasing classification performance from CLAM to Transformer to GNN.
On the OOD dataset, the Transformer with the feature extractor pre-trained on ImageNet leads to the best performance.
In contrast, the GNN with Ciga feature extractor performs quite poorly on the OOD data.
Except for the GNN-Ciga case, we observe that the Transformer and GNN perform better on ID and OOD data compared to the CLAM baseline.
For the evaluations with respect to ECE, we mostly see overlapping intervals between the results for the five model checkpoints across the different aggregation methods. 
We observe a trend that the models using the Ciga feature extractor appear to be slightly better calibrated than the ones using the feature extractor pre-trained on ImageNet.
Especially for the OOD data, we see that CLAM and GNN with the ImageNet feature extractor lead to the worst ECE values.

\begin{figure}[htbp]
    \includegraphics[width=\textwidth]{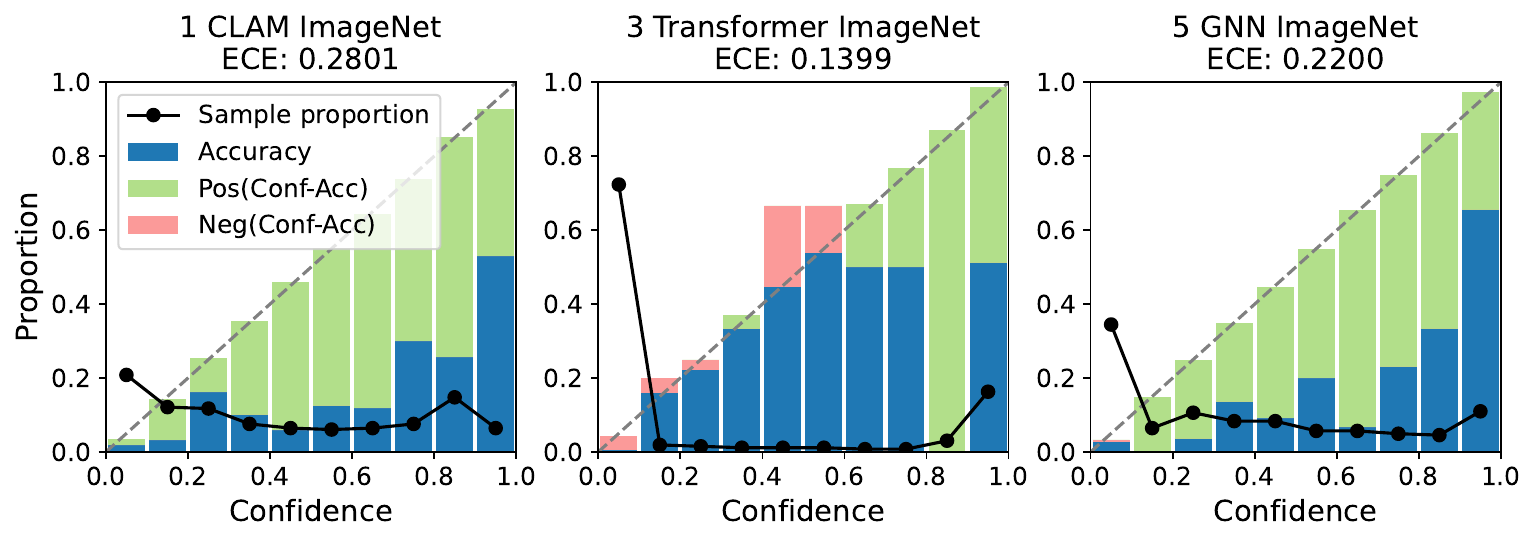}
    \caption{Reliability diagrams for the three aggregation methods with ImageNet feature extractor.}
    \label{fig:reliability}
\end{figure}

When looking at the aggregated ECE metric, we do not observe notable differences between the three included aggregation methods.
As an additional step, in \autoref{fig:reliability} we show an exemplary reliability diagram for each aggregation method with an ImageNet feature extractor evaluated on the ID test set.
The reliability diagrams show the average accuracy for each confidence bin and we additionally overlay the sample proportion for each bin, as the ECE is computed accordingly.
Since we evaluate a binary classification problem, we look at the confidence values for the positive MSI class.
When looking at the last bin with confidence values ranging from $0.9$ to $1.0$, we see that only $50 \%$ to $70 \%$ of the predictions are correct across all three methods.
For a well-calibrated model, the accuracy in this bin should be above $90 \%$.
An additional observation we make is that the Transformer architecture appears to push the output probabilities to extreme values of either $0.0$ or $1.0$. 
This behavior is not reflected in the ECE value and can only be observed when looking at the reliability diagrams.

To improve the calibration of the three aggregation methods, we apply Temperature Scaling (c.f. Section \ref{sec:calibration-and-ts}) to the output logits.
To compute the optimal temperature $T$ for each method, we optimize the Negative Log-Likelihood (NLL) w.r.t. $T$ on the validation set for each fold.
\autoref{tab:ts_results} shows the impact of Temperature Scaling on the calibration error for both ID and OOD test data.
When looking at the table, we do not observe notable improvement in ECE when applying Temperature Scaling.
For some ID cases (CLAM Ciga, GNN ImageNet, GNN Ciga) the mean ECE even slightly increases when applying Temperature Scaling.

\begin{table}[htbp]
    \centering
    \caption{Impact of Temperature Scaling (TS) on calibration error}
    \label{tab:ts_results}
    \begin{tabular}{lllll}
    Method                  & ID ECE & ID ECE (TS) & OOD ECE & OOD ECE (TS) \\
    \hline
    CLAM ImageNet           
    & $0.251 \pm 0.048$ & $\mathbf{0.248 \pm 0.051}$ & $0.325 \pm 0.066$ & $\mathbf{0.319 \pm 0.066}$ \\
    CLAM Ciga               
    & $\mathbf{0.199 \pm 0.041}$ & $0.202 \pm 0.039$ & $0.167 \pm 0.047$ & $\mathbf{0.163 \pm 0.048}$ \\
    Transformer ImageNet    
    & $0.144 \pm 0.033$ & $\mathbf{0.136 \pm 0.035}$ & $0.206 \pm 0.086$ & $\mathbf{0.197 \pm 0.089}$ \\
    Transformer Ciga        
    & $0.164 \pm 0.014$ & $\mathbf{0.158 \pm 0.010}$ & $0.160 \pm 0.015$ & $\mathbf{0.149 \pm 0.019}$ \\
    GNN ImageNet            
    & $\mathbf{0.192 \pm 0.048}$ & $0.195 \pm 0.053$ & $0.310 \pm 0.087$ & $\mathbf{0.306 \pm 0.087}$ \\
    GNN Ciga                
    & $\mathbf{0.152 \pm 0.060}$ & $0.155 \pm 0.064$ & $0.150 \pm 0.035$ & $\mathbf{0.144 \pm 0.040}$ \\
    \hline
    \end{tabular}
\end{table}

\section{Discussion and Conclusion}
\label{sec:discussion-conclusion}

Within this work, we have evaluated three feature aggregation methods with respect to their classification performance and calibration for the task of predicting MSI from histological images.
When comparing different feature extractors and different magnification levels, we observe that the ResNet pre-trained on histological images \cite{ciga2022} does not lead to notable improvement in performance compared to the ResNet pre-trained on ImageNet \cite{deng2009}.
However for the MSI classification task we can note that using image patches with 20x magnification is sufficient and even leads to better performance compared to using image patches at 40x magnification.

When comparing the AUROC values for the three feature aggregation methods, we observe that both the Transformer as well as the GNN architecture perform better than the CLAM baseline on the ID test data.
On the OOD test data, the Transformer with ImageNet features performs best.
Concerning calibration in terms of ECE, we observe similar results across methods.
While CLAM ImageNet and GNN ImageNet have higher ECE values on OOD data, we can say that Transformers lead to equally good calibration results on ID and OOD data.
These findings are in line with \cite{minderer2021}, who also observe good performance of Transformers in terms of generalization and calibration.
An opportunity to improve the applied Transformer architecture would be to avoid limiting the number of patches due to computational constraints.
Possibilities include the adjustment of the attention mechanism as proposed for example in \cite{wang2020c, beltagy2020} to make the computational complexity scale linearly with the sequence length.

When we look at exemplary reliability diagrams for each of the three aggregation methods, the promising ECE values of the Transformer architecture have to be taken with a grain of salt.
For Transformers, we notice that the output probabilities are pushed to the extreme values of either $0.0$ or $1.0$.
While it is fine for a perfect classifier to identify all classes with $100 \%$ confidence, in the medical context we are especially concerned about wrong predictions with high confidences.
Since training a classifier with hard labels is designed to lead to these extreme output values, it would be worth to consider the integration of soft labels.
Especially in the medical context, diagnoses are often not $100 \%$ certain and often vary between observers. 
Therefore it would be worth to generate an average label from multiple observers and train the network with the soft label.
This would in the end lead to more meaningful output confidences further away from the extrema.
However, for the current work we were limited to one label per sample and therefore could not investigate the usage of soft labels.

As method to improve calibration, we applied Temperature Scaling to the output logits of the three aggregation methods.
In contrast to previous works \cite{guo2017, minderer2021} we could not observe improvements in terms of ECE when applying Temperature Scaling.
This result might be caused by the comparably small amount of samples in general and therefore also the small amount of samples in the validation set.
As we could not configure Temperature Scaling to lead to consistent improvements in calibration error, it would be worth trying other methods in future works.
One option would be to train an ensemble of methods, as deep ensembles have been shown to improve ECE on classical image datasets \cite{ovadia2019}.
As the usage of deep ensembles leads to a high increase in computational requirements, it would be more favorable to find solutions that improve calibration for the output probabilities of a single network.

From our evaluations, we can say that both Transformers and GNNs are promising network architectures for slide-level WSI classification and it will be worth to develop new methods to improve the calibration of these methods.

\section*{Acknowledgements}
The research is funded by the \textit{Ministry of Social Affairs, Health and Integration} of the Federal State Baden-Württemberg, Germany.

\bibliographystyle{plainnat} 
\bibliography{main}  






\end{document}